\documentclass[journal]{IEEEtran}
\usepackage{cite}
\usepackage{amsmath,amssymb,amsfonts}
\usepackage{algorithmic}
\usepackage{graphicx,graphics}
\usepackage{textcomp}
\usepackage{xcolor}
\usepackage{algorithm}
\usepackage{multirow}
\usepackage[acronym]{glossaries}
\usepackage{caption}
\usepackage{subcaption}
\usepackage{bbm}
\usepackage{breqn}
\usepackage{dsfont}

\DeclareMathOperator*{\argmaxB}{argmax}

\title{Rank-Aware Link Adaptation for XR Tethering Groups with Realistic Tethering Link: A Multi-Offset OLLA Framework  
}

\makeglossaries
\newacronym{5g}{5G}{5th generation}
\newacronym{5ga}{5G-A}{5th Generation-Advanced}
\newacronym{3gpp}{3GPP}{3rd Generation Partnership Project}
\newacronym{ack}{ACK}{acknowledgment}
\newacronym{ar}{AR}{augmented reality}
\newacronym{ap}{AP}{Access Point}
\newacronym{ampdu}{A-MPDU}{aggregate-MAC protocol data unit}
\newacronym{awgn}{AWGN}{additive white Gaussian noise}
\newacronym{bler}{BLER}{block error rate}
\newacronym{blep}{BLEP}{block error probability}
\newacronym{cb}{CB}{code block}
\newacronym{cbg}{CBG}{code block group}
\newacronym{cts}{CTS}{clear to send}
\newacronym{cc}{CC}{Chase combining}
\newacronym{cdf}{CDF}{cumulative distribution function}
\newacronym{csi}{CSI}{channel state information}
\newacronym{csi-rs}{CSI-RS}{channel state information reference signal}
\newacronym{cqi}{CQI}{channel quality indicator}
\newacronym{dci}{DCI}{downlink control information}
\newacronym{dpp}{DPP}{Drift-Plus-Penalty}
\newacronym{du}{DU}{Dense Urban}
\newacronym{dl}{DL}{downlink}
\newacronym{dlsch}{DL-SCH}{downlink shared channel}
\newacronym{dmrs}{DMRS}{demodulation reference signals}
\newacronym{df}{DF}{decode-and-forward}
\newacronym{embb}{eMBB}{enhanced mobile broadband}
\newacronym{ecdf}{eCDF}{empirical cumulative distribution function}
\newacronym{fdd}{FDD}{Frequency Division Duplexing}
\newacronym{fb}{FB}{feedback}
\newacronym{gNodeB}{gNodeB}{Next-Generation Node B}
\newacronym{harq}{HARQ}{hybrid automatic repeat request}
\newacronym{illa}{ILLA}{Inner Loop Link Adaptation}
\newacronym{inh}{InH}{Indoor Hotspot}
\newacronym{ici}{ICI}{inter-cell interference}
\newacronym{jhra}{JHRA}{Joint HARQ Retransmission Algorithm}
\newacronym{la}{LA}{Link Adaptation}
\newacronym{los}{LOS}{line-of-sight}
\newacronym{ldpc}{LDPC}{Low Density Parity Check}
\newacronym{llr}{LLRs}{log-likelihood ratios}
\newacronym{lrb}{LRBs}{least reliable soft bits}
\newacronym{mac}{MAC}{medium access control}
\newacronym{map}{MAP}{maximum a posteriori}
\newacronym{mpdu}{MPDUs}{MAC protocol data units}
\newacronym{mmib}{MMIB}{mean mutual information per bit}
\newacronym{mmwave}{mmWave}{millimeter-wave}
\newacronym{mrb}{MRBs}{most reliable soft bits}
\newacronym{mimo}{MIMO}{Multiple-Input and Multiple-Output}
\newacronym{mbs}{MBS}{multicast and broadcast services}
\newacronym{mmseirc}{MMSE-IRC}{minimum mean square error-interference rejection combining}
\newacronym{mcs}{MCS}{modulation and coding scheme}
\newacronym{moolla}{MO-OLLA}{multi-offset Outer Loop Link Adaptation}
\newacronym{mi}{MI}{mutual information}
\newacronym{mr}{MR}{mixed reality}
\newacronym{nr}{NR}{new radio}
\newacronym{nack}{NACK}{negative acknowledgement}
\newacronym{nlos}{NLOS}{non-line-of-sight}
\newacronym{nas}{NAS}{Non-Access Stratum}
\newacronym{ofdm}{OFDM}{orthogonal frequency-division multiplexing}
\newacronym{olla}{OLLA}{Outer Loop Link Adaptation}
\newacronym{pdb}{PDB}{packet delay budget}
\newacronym{pmi}{PMI}{precoding matrix indicator}
\newacronym{ptm}{PTM}{point-to-multipoint}
\newacronym{ptp}{PTP}{point-to-point}
\newacronym{prb}{PRBs}{Physical Resource Blocks}
\newacronym{pdsch}{PDSCH}{physical downlink shared channel}
\newacronym{pdcch}{PDCCH}{physical downlink control channel}
\newacronym{quadriga}{QuaDRiGa}{QUAsi Deterministic RadIo channel GenerAtor}
\newacronym{qos}{QoS}{quality of service}
\newacronym{re}{RE}{resource element}
\newacronym{rts}{RTS}{request to send}
\newacronym{rrm}{RRM}{radio resource management}
\newacronym{rx}{Rx}{Receive}
\newacronym{rrc}{RRC}{Radio Resource Control}
\newacronym{sc}{SC}{selection combining}
\newacronym{softc}{SoftC}{soft combining}
\newacronym{scs}{SCS}{Selection Combining Scheme}
\newacronym{sscs}{SSCS}{Selection/Soft Combining Scheme}
\newacronym{svd}{SVD}{singular value decomposition}
\newacronym{s-c-s}{S-C-S}{sub-carrier spacing}
\newacronym{sls}{SLS}{system-level simulator}
\newacronym{softdf}{soft-DF}{soft decode-and-forward}
\newacronym{siso}{SISO}{soft-input soft-output}
\newacronym{sinr}{SINR}{signal-to-interference-plus-noise ratio}
\newacronym{snr}{SNR}{signal-to-noise ratio}
\newacronym{srs}{SRS}{sounding reference signal}
\newacronym{saw}{SAW}{stop-and-wait}
\newacronym{thgr}{TGr}{tethering group}
\newacronym{tl}{TL}{tethering link}
\newacronym{tb}{TB}{transport block}
\newacronym{tbler}{TBLER}{transport block error rate}
\newacronym{tx}{Tx}{Transmit}
\newacronym{tdd}{TDD}{Time Division Duplex}
\newacronym{uci}{UCI}{uplink control information}
\newacronym{ue}{UE}{User Equipment}
\newacronym{ue-x}{UE-X}{User Equipment-extended reality}
\newacronym{ue-t}{UE-T}{User Equipment with Tethering}
\newacronym{ue-k}{UE-$k$}{User Equipment k}
\newacronym{ul}{UL}{Uplink}
\newacronym{urllc}{URLLC}{Ultra Reliable Low Latency Communications}
\newacronym{vr}{VR}{virtual reality}
\newacronym{vrb}{VRB}{Virtual Resource Block}
\newacronym{wlan}{WLAN}{wireless local-area network}
\newacronym{wifi}{WiFi}{Wireless Fidelity}
\newacronym{xr}{XR}{Extended Reality}

\author{Muhammad Ahsen, \IEEEmembership{Graduate Student Member, IEEE}, Boyan Yanakiev, Claudio Rosa, and Ramoni Adeogun, \IEEEmembership{Senior Member, IEEE}

\thanks{Muhammad Ahsen and Ramoni Adeogun are with the Department of Electronic Systems, Technical Faculty of IT and Design, Aalborg University, 9220 Aalborg, Denmark (e-mail: muah@es.aau.dk; ra@es.aau.dk).} 
\thanks{Boyan Yanakiev and Claudio Rosa are with the Nokia, DK-9220 Aalborg, Denmark (e-mail:boyan.yanakiev@nokia.com; claudio.rosa@nokia.com)}}

\begin{document}
\newcommand{\roa}[1]{\textcolor{blue}{[Ramoni: #1]}}

\maketitle

\begin{abstract}
We investigate higher-rank transmissions for multi-connected \gls{xr} devices enabled through \gls{thgr}, in which a nearby tethering \gls{ue} cooperates with an \gls{xr} \gls{ue} via a short-range \gls{tl}.
In contrast to prior studies that are limited to rank-1 transmission and ideal tethering assumptions, we analyze \gls{thgr} performance under higher-rank \gls{ptm} transmission and realistic \gls{tl} delays.
Conventional single \gls{olla} offset results in inaccurate throughput prediction across ranks, leading to suboptimal rank selection.
To address this limitation, we propose a \gls{moolla} framework that introduces rank-dependent \gls{sinr} correction to improve \gls{la} accuracy.
Furthermore, a \gls{wifi} based delay model is incorporated to characterize the impact of practical \gls{tl} constraints including limited bandwidth and achievable throughput on \gls{xr} capacity and cellular resource utilization, providing the first such analysis for higher-rank multi-connected \gls{xr} device.
System-level simulations demonstrate that \gls{moolla} provides up to 20\% performance improvement over conventional \gls{olla} for multi-connected \gls{xr} \gls{ue}s. 
Moreover, \gls{thgr}s effectively exploit higher-rank transmission, achieving \gls{xr} capacity gains of 180-200\% over single-link \gls{xr} \gls{ue}s under ideal \gls{tl} conditions.
Critically, the gains of the \gls{thgr} remain at 165-180\% under realistic high-throughput \gls{tl}s relative to single-link \gls{xr} \gls{ue}s, confirming the practical viability of \gls{thgr} based cooperation for \gls{xr} capacity enhancements within existing cellular resources.

\end{abstract}


\section{Introduction}

\gls{xr} encompassing \gls{vr}, \gls{ar} and \gls{mr} is considered as a defining application for next-generation wireless systems, imposing stringent and simultaneous demands on latency, reliability and throughput that far exceed those of conventional mobile broadband services \cite{bib1}.
Meeting these demands within the constrained radio resources of a cellular network significantly limits the number of \gls{xr} users that can be simultaneously supported within a \gls{5ga} cellular cell \cite{bib3-1,bibr4}.
To support a higher \gls{xr} users, not only link-level reliability is required but also higher spectral efficiency, making spatial multiplexing through higher-rank transmission a natural and necessary tool for increasing the number of simultaneously supportable XR users. 
However, realizing spatial multiplexing gains in \gls{xr} scenarios is non-trivial, the stringent latency and reliability constraints imposed by \gls{xr} \gls{qos} requirements demand accurate link adaptation.
Moreover, the interference suppression capability of conventional receivers such as \gls{mmseirc} can degrade under higher-rank transmissions \cite{bib7-14} particularly in \gls{ici} limited environments. 
These challenges motivate the investigation of transmission schemes that can exploit spatial multiplexing while preserving the decoding reliability required for \gls{xr} services.

Multi-connectivity has been explored as a means to simultaneously improve both reliability \cite{bib4-1, bib8} and spectral efficiency \cite{bib2-our2, bib2-our4} for \gls{xr}. 
Among emerging approaches, \gls{xr} \gls{thgr}, in which an \gls{xr} \gls{ue} is cooperatively paired with a nearby conventional \gls{ue} via a short-range \gls{tl}, have demonstrated compelling capacity gains using only sub-6 GHz resources, without requiring additional cellular spectrum \cite{bib2-our2, bib2-our4}. 
The cooperative reception mechanism within a \gls{thgr}, combining joint \gls{harq} processing, soft combining, and joint \gls{olla}, improves post-combining \gls{sinr} and reduces effective \gls{bler} for a given channel condition. 
Crucially, this enhanced decoding reliability not only improves link robustness but also creates more favorable conditions for higher-rank transmission, enabling \gls{thgr} to operate at higher \gls{mcs} and rank than would be feasible for single-link legacy \gls{ue}s under the same interference conditions. 
Whether and to what extent \gls{thgr}s can exploit this advantage to achieve higher \gls{xr} capacity through spatial multiplexing remains an open and practically important question.

\gls{la} plays a critical role in realizing these gains by selecting appropriate transmission parameters based on channel conditions \cite{bib3-1,bib6-1}.
Existing studies on \gls{xr} \gls{thgr} have primarily focused on rank-1 transmission and ideal \gls{tl}s \cite{bib2-our2,bib2-our4}, where a joint \gls{olla} mechanism was shown to improve \gls{mcs} selection within \gls{thgr}, effectively translating cooperation gains into spectral efficiency improvements.
However, the performance of \gls{thgr}s under higher-rank transmission remains largely unexplored.

The authors of \cite{bib3-6} studied \gls{xr} capacity for legacy \gls{ue}s under higher-rank (rank-2) transmissions but did not compare their results against rank-1 scenarios.
Moreover, higher-rank transmission can degrade interference suppression capability of the \gls{mmseirc} receiver \cite{bib7-14}.
For legacy \gls{ue}s, this degradation in interference suppression can outweigh the spatial multiplexing gains, leading to suboptimal or even negative capacity outcomes from higher-ranks.
For \gls{thgr}s, however, the joint \gls{olla} and \gls{harq} algorithms enable operation at higher spectral efficiency, which in turn reduces \gls{ici} and may mitigate this limitation.
Nevertheless, the net impact of higher-rank transmission on \gls{thgr} performance, particularly in terms of \gls{xr} capacity, cannot be assumed a priori, especially in \gls{ici} limited environments such as \gls{inh} deployments.

Furthermore, conventional rank adaptation methods are typically based on greedy throughput maximization \cite{bib7-15} and do not explicitly account for \gls{olla} \cite{bib7-16}.
Even when \gls{olla} is incorporated, a single-offset is applied uniformly across all ranks.
This is fundamentally problematic because different transmission ranks operate at different \gls{sinr} regimes and exhibit different effective \gls{bler} characteristics.
A single-offset leads to throughput prediction errors and suboptimal rank and \gls{mcs} selection, as different ranks exhibit distinct \gls{sinr} operating regimes.
For \gls{xr} services, where such inefficiency directly reduces the number of supportable users per cell, this limitation has significant practical consequences. 

Beyond \gls{la}, the practical deployment of \gls{thgr} also raises important questions about the \gls{tl}.
When cooperation is utilized, the \gls{tl} delay may impact the \gls{saw} channel utilization of the network and more importantly increases the application layer delay, which may affect the performance gains of the \gls{thgr}.
Whereas, the delay over the \gls{tl} largely depends on the \gls{tl} throughput and its load.
Despite its practical relevance, the impact of realistic \gls{tl} constraints, including limited bandwidth and delay, on \gls{thgr} performance has not been previously characterized.


To address these gaps, this paper investigates higher-rank transmission for \gls{xr} \gls{thgr} in the \gls{inh} scenario and analyzes the impact of realistic \gls{tl} delays.
In addition, we propose an \gls{moolla} framework that enables rank aware \gls{sinr} correction for improved \gls{la}.

The main contributions of this paper are summarized as follows
\begin{itemize}
    \item We investigate the performance of \gls{xr} \gls{thgr}s under higher-rank \gls{ptm} transmission and quantify the resulting \gls{xr} capacity gains relative to legacy \gls{xr} \gls{ue}s. The results demonstrate that \gls{thgr}s can effectively exploit spatial multiplexing to achieve substantially higher \gls{xr} capacity than is attainable by single-link \gls{ue}s with rank adaptation, an advantage rooted in the \gls{thgr}'s ability to overcome the spatial multiplexing vs interference trade-off that limits conventional \gls{ue}s. 
    \item We propose \gls{moolla}, a rank aware \gls{la} scheme that maintains independent \gls{olla} offsets per rank to enable more accurate \gls{sinr} estimation and throughput prediction. 
    \item We model the \gls{tl} using a \gls{wifi} based delay abstraction and conduct the first systematic analysis of how practical \gls{tl} constraints, including limited PHY rate and queuing delay, interact with network's \gls{harq} resources and \gls{xr} capacity, establishing a feasibility boundary for \gls{thgr} deployment under realistic tethering conditions.  
\end{itemize}

The remainder of the paper is organized as follows.
Section \ref{systemmodel} presents the system model, including the cooperative reception framework and the \gls{tl} delay model.
Section \ref{aOLLA} describes the \gls{la} framework and presents the proposed \gls{moolla} scheme.
Section \ref{simrst} provides the performance evaluation methodology and discusses the results.
Section \ref{conc} concludes the paper.

\section{System Model}
\label{systemmodel}

\begin{figure}
\centerline{\includegraphics[width=8cm]{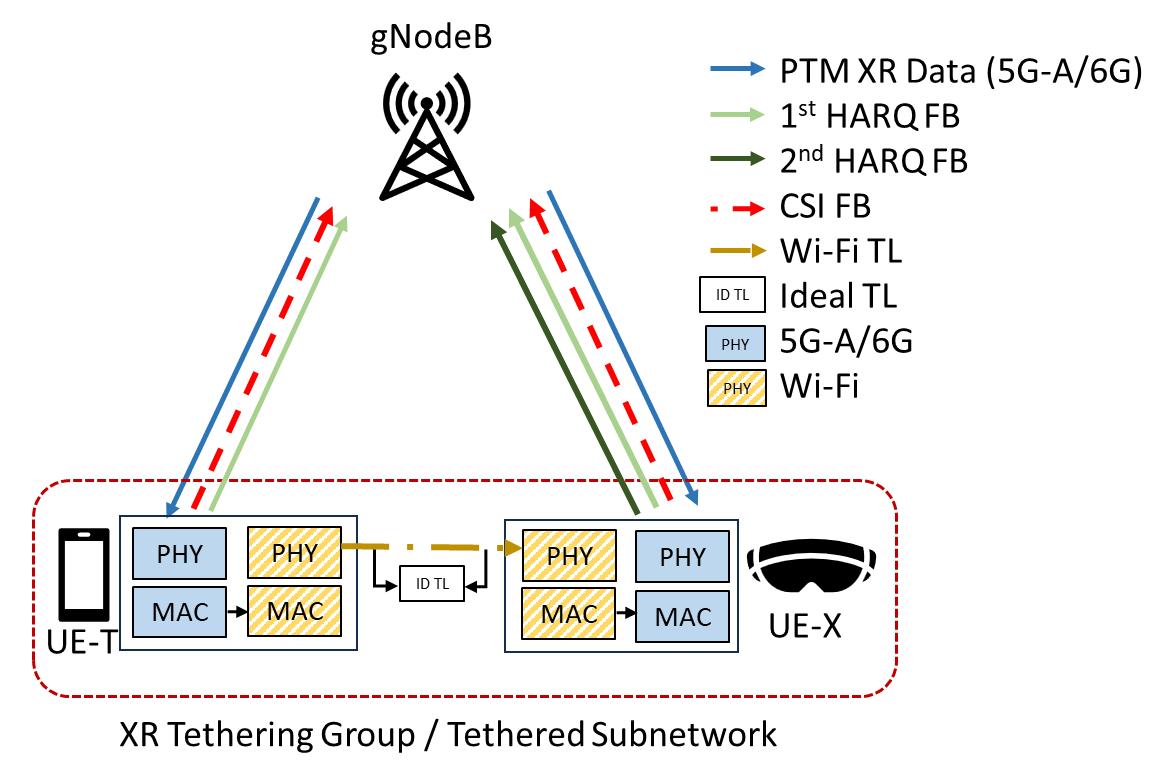}}
\caption{An illustration of a single \gls{thgr} connected to a \gls{gNodeB}}
\label{fig:TGr}
\end{figure}

We consider a \gls{dl} cellular network comprising multiple cells, where each cell serves multiple \gls{xr} \gls{thgr}s with a uniform load distribution, i.e., the same number of \gls{xr} \gls{thgr}s is connected to each \gls{gNodeB}.
Each \gls{xr} \gls{thgr} consists of two closely located \gls{ue}s, an \gls{xr} \gls{ue} (\acrshort{ue-x}) and a conventional \gls{ue} (\acrshort{ue-t}), as illustrated in Fig. \ref{fig:TGr}.
Both \gls{ue}s are associated with the same \gls{gNodeB}.
In addition, the two \gls{ue}s are connected to each other via a short-range \gls{tl}, which may be implemented using a \acrlong{wlan} interface or sidelink \cite{bib6}.

The network configures both \gls{ue}s of the \gls{thgr} into a tethered-assisted session.
The tethered-assisted session is conceptually analogous to a \gls{5g} multicast session \cite{bib11}.
In this session, \gls{dl} data is delivered using \gls{ptm} transmission, in which the same \gls{tb} is transmitted to both \gls{ue}s of the \gls{thgr} using a single set of cellular radio resources. 
\gls{ul} transmissions remain \gls{ue} specific and are independently used for \gls{harq} feedback and \gls{csi} reporting, in accordance with standard specifications.

Asynchronous \gls{harq} is assumed and both \gls{ue}s in a \gls{thgr} provide \gls{harq} feedback prior to any cooperation, relative to the initial \gls{tb} decoding attempt \cite{bib12}.
This initial \gls{harq} feedback is unaffected by the \gls{tl} delay.
However, \acrshort{ue-x} may provide an additional, second \gls{harq} feedback when soft combining is applied just like a \gls{ue} does when it receives a retransmission from \gls{gNodeB}.
This second feedback is affected by the \gls{tl} delay and may delay retransmissions from the \gls{gNodeB} if \acrshort{ue-x} fails to decode the \gls{tb} even after soft combining and sends a \gls{nack}.
\gls{harq} feedback from both \gls{ue}s is processed using a joint \gls{harq} processing algorithm to support retransmissions within the tethered-assisted session \cite{bib2-our2}.
According to this algorithm, the \gls{gNodeB} retransmits only if the \gls{thgr} fails to decode the \gls{tb} even after cooperation, depending on the cooperation scheme employed.
The reported \gls{csi}, together with the \gls{harq} feedback from both \gls{ue}s, is utilized at the \gls{gNodeB} for link adaptation via a joint \gls{olla} algorithm \cite{bib2-our4}.
Alternatively, the second \gls{harq} feedback can be modeled as delayed \gls{harq} feedback from \acrshort{ue-x}, in which \acrshort{ue-x} transmits only a single feedback message, either an \gls{ack} upon immediate successful decoding or a delayed feedback after cooperation if initial decoding fails. 


\subsection{Cooperative Reception Model}
Each \gls{xr} \gls{thgr} operates according to a \gls{sscs} \cite{bib2-our4}.
Both \gls{ue}s initially attempt to decode the received \gls{tb}.
Depending on the decoding outcomes at \acrshort{ue-x} and \acrshort{ue-t}, multiple scenarios may arise, as detailed in \cite{bib2-our4}.
Ultimately, the \gls{thgr} decodes the \gls{tb} according to one of the following three scenarios. 
\begin{enumerate}
    \item Decoding at \acrshort{ue-x} only ($D_{X}$): If \gls{ue-x} successfully decodes the \gls{tb}, it accepts the decoded data and any corresponding information forwarded by \gls{ue-t} is discarded. This scenario incurs no \gls{tl} delay. 
    \item Decoding after \acrlong{sc} ($D_{SC}$): If \gls{ue-x} fails to decode the \gls{tb} but \gls{ue-t} succeeds, \acrshort{ue-t} forwards the decoded \gls{tb} over the \gls{tl} to \gls{ue-x} using a decode-and-forward algorithm \cite{bib18-2}, which applies selection combining. In this case, the \gls{tl} delay does affect the \gls{tb} delay and is calculated using the decoded \gls{tb}. 
    \item Decoding after soft combining ($D_{SoftC}$): If both \gls{ue}s fail to decode the \gls{pdsch} but successfully decode the \gls{pdcch}, \acrshort{ue-t} forwards the soft-\gls{tb} (\gls{llr} \cite{bib7-20} corresponding to that \gls{tb}) to \acrshort{ue-x} \cite{bib2-1}. \acrshort{ue-x} then performs element-wise soft combining \cite{bib2-3} followed by a second decoding attempt. This scenario also incurs an additional delay approximately equal to the \gls{tl} delay of the quantized soft-\gls{tb}.   
\end{enumerate}

\subsection{Tethering Link}
\label{systemmodeltl}
The \gls{tl} is assumed to operate on a carrier frequency different from that used by the cellular network for \gls{dl} and \gls{ul} transmissions.
It may utilize \gls{wifi} or sidelink \cite{bib6} or any other proprietary solution. 

\subsubsection{Delay over Tethering Link}
In this work, the \gls{tl} is first assumed to be ideal, with infinite bandwidth and zero delay, to establish upper-bound \gls{thgr} gains.
Subsequently, the analysis is repeated with the \gls{tl} nodeled using \gls{wifi} technology.  
When operating over \gls{wifi}, decoded \gls{tb}s or \gls{llr} are forwarded by aggregating multiple \gls{mpdu} as \gls{ampdu}.
The total delay $D_{n}$ associated with transmitting data $n$ (i.e., decoded \gls{tb} or \gls{llr}) over the \gls{tl} is given by

\begin{equation}
    D_n = W_n + S_n
\end{equation}
where $W_n$ denotes the waiting time in the queue prior to transmission and $S_n$ denotes the service time required for successful transmission.

The arrival of \gls{tb}s (or \gls{llr}) at the \gls{wifi} interface depends on cellular network conditions, such as scheduling decisions, channel quality and the number of active users in the system.
In contrast, the service time depends on \gls{wifi} parameters including channel bandwidth, collision probability and \gls{mcs}.
Since both the inter-arrival and service time distributions are non-exponential in \gls{thgr} (cellular/\gls{wifi} systems), $W_n$ is modeled using a G/G/1 queue with general arrival and general service distributions \cite{bib7-19} and is computed using Lindley's recursion \cite{bib7-13}

\begin{equation}
\label{eq:w}
    W_{n+1} = \max (0, W_n +S_n-A_n)
\end{equation}

where $A_n$ represents the inter-arrival time between consecutive \gls{tb}s of the user.

Given the short distance between \acrshort{ue-t} and \acrshort{ue-x}, the \gls{tl} is assumed to experience negligible packets errors,
however, collisions may still occur.
Let $p_c$ denote the collision probability.
The service time $S_n$, which captures the stochastic behavior of \gls{wifi} follows the IEEE 802.11 distributed coordination function(DCF) analysis framework \cite{bib7-5,bib7-4}, where the total service time accounts for backoff stages, collision durations and successful \gls{ampdu} transmission   

\begin{equation}
    S_n = \sum_{i=0}^{N_n} B_{n,i} + N_n . T_c + T_{s,n} 
\end{equation}

where $N_n$ is the number of collisions before successful transmission (bounded by $N_{max}$), $B_{n,i}$ is the backoff time before the $i$-th transmission attempt, $T_c$ is the collision duration and $T_{s,n}$ is the physical transmission time of the \gls{ampdu}.
The number of failed transmission attempts is modeled as Bernoulli trials, where each attempt succeeds with probability ($1-p_c$).
The backoff time $B_{n,i}$ follows the IEEE 802.11 DCF binary exponential backoff mechanism \cite{bib7-5}
\begin{equation}
    B_{n,i} = \mathcal{U}(0,CW_i-1) \times \sigma
\end{equation}
where $CW_{i} = \min (2^i . CW_{min}, CW_{max})$ is the contention window at stage $i$ with $CW_{min}$ and $CW_{max}$ being the minimum and maximum contention windows, respectively. 
$\sigma$ is the slot duration.

For an \gls{ampdu}, the transmission time $T_{s,n}$ can be calculated as \cite{bib7-12}  
\begin{equation}
    T_{s,n} = T_{RTS}+3 \times SIFS+T_{CTS} + T_{Data_n} + T_{ACK} + DIFS + \sigma
\end{equation}
where $T_{RTS}$, $T_{CTS}$, $SIFS$, $DIFS$ and  $\sigma$ are standard parameters of \gls{wifi} \cite{bib7-5}.
The data transmission time is approximated as \cite{bib7-12} 
\begin{equation}
    T_{Data_n} \approx T_{PHY} + \lceil\frac{L_H + L_n}{DBPS} \rceil T_{OFDM}
\end{equation}
where $T_{PHY}$ is the preamble duration, $L_H$ is the \gls{mac} header, $L_n$ is the \gls{ampdu} payload size for data $n$, $DBPS$ denotes the number of data bits per \gls{ofdm} symbol (dependent on the number of subcarriers, channel bandwidth, number of spatial streams, \gls{mcs} and code rate) and $T_{OFDM}$ is the \gls{ofdm} symbol duration.
The ratio $DBPS/T_{OFDM}$ defines the PHY rate of the link.
The collision duration $T_c$ is given by \cite{bib7-4} 
\begin{equation}
    T_c = T_{RTS}+ SIFS+T_{CTS}  + DIFS + \sigma
\end{equation}
If a decoded \gls{tb} is forwarded then $L_n$ corresponds to the \gls{tb} size.
If \gls{llr} are forwarded, they are first quantized, accordingly $L_n$ corresponds to the number of the \gls{llr} multiplied by the number of quantization bits, where the quantization precision is chosen as the minimum number of bits required to represent each LLR value.

\section{\acrlong{la} for \gls{thgr}}
\label{aOLLA}
In conventional rank adaptation, the \gls{sinr}s $\gamma^{r}$ corresponding to each rank $r$, obtained via \gls{csi-rs} or \gls{srs}, are used to select the rank $r^*$ that maximizes the achievable throughput $R$ with the highest feasible \gls{mcs} index $m*$ for the \gls{ue}, as given by
\begin{equation}
    r^* = \argmaxB_{r} R [\gamma^r,m^*]
\end{equation}
An \gls{olla} correction is applied to $\gamma^{r}$ to obtain the effective \gls{sinr} $\gamma_{eff}^{r}$.

\begin{equation}
\label{eq:sinr}
    \gamma_{eff}^{r} = \gamma^{r} - \Delta_{OLLA}, \forall r \in \{1,2,.. max \}
\end{equation}
where $\Delta_{OLLA}$ is the \gls{olla} offset, adjusted based on the \gls{harq} feedback and step sizes corresponding to the target \gls{tbler} $TBLER^T$, compensating for \gls{cqi} inaccuracies and/or reporting delays for legacy \gls{ue}s.
For \gls{thgr}, the \gls{olla} offset additionally captures the cooperation gains achieved through the joint \gls{olla} algorithm \cite{bib2-our4}.
The use of the effective \gls{sinr} from \eqref{eq:sinr} in rank and \gls{mcs} selection, effectively results in rank and \gls{mcs} selection based on $TBLER^T$: 

\begin{multline}
    \label{eq:rankb}
    (r^*, m^*) = \argmaxB_{r,m} [ R \left( \gamma_{eff}^{r}, m \right) | \\
    \left( 1- P^{succ}(\gamma_{eff}^{r},m) \right) \leq {TBLER}^{T}  ]        
\end{multline}

where $P^{succ}(\gamma_{eff}^{r},m)$ is the perceived probability of successful \gls{tb} decoding at rank $r$ with \gls{mcs} $m$.
This rank and \gls{mcs} selection procedure serves as the baseline for both legacy \gls{ue}s and \gls{thgr}s.
For \gls{thgr}, the \gls{sinr} $\gamma^{r}$ depends on the \gls{csi} reporting scheme, i.e., whether both \gls{ue}s report \gls{csi} and the best is selected or only \acrshort{ue-x} reports \gls{csi} \cite{bib2-our4}.
Accordingly, $\gamma^{r}$ for \gls{thgr} is given as
\begin{equation}
    \gamma^{r} = \begin{cases}
        \max \left( \gamma_{\acrshort{ue-x}}^r, \gamma_{\acrshort{ue-t}}^r\right), & \text{if \gls{csi} Best} \\
        \gamma_{\acrshort{ue-x}}^r, & \text{if \gls{csi} \acrshort{ue-x}} \\
    \end{cases}
\end{equation}
Similarly, the \gls{olla} offset for a \gls{thgr} is updated based on joint \gls{harq} feedback $HF_J$, which incorporates feedback from both \gls{ue}s as well as the optional delayed feedback from \acrshort{ue-x} following soft combining \cite{bib2-our2}.
The joint \gls{harq} feedback $HF_J$ is defined as
\begin{equation}
    HF_J = HF_{X1} \vee HF_{T1} \vee HF_{X2}
\end{equation}
 where $HF_{X1}$ and $HF_{T1}$ denote the first \gls{harq} feedback provided by \acrshort{ue-x} and \acrshort{ue-t}, respectively, following the initial \gls{tb} decoding attempt and $HF_{X2}$ represents the conditional second (delayed) \gls{harq} feedback from \acrshort{ue-x} following soft combining.
Further details are available in our prior work \cite{bib2-our2}.

Applying a single \gls{olla} offset uniformly across all ranks is suboptimal, as the \gls{sinr} associated with each rank requires different degrees of correction to achieve the $TBLER^T$.
This mismatch produces a cross-rank contamination in throughput prediction, which in turn results in suboptimal rank and \gls{mcs} selection.


The fundamental limitation of single-offset \gls{olla} under multi-rank operation is best understood by examining the two distinct timescales on which the offset and rank selection operate. 
The \gls{olla} offset $\Delta_{OLLA}$ is updated at the fast timescale of \gls{harq} feedback, after every \gls{tb} transmission, the \gls{harq} outcome (\gls{ack} or \gls{nack}) updates the offset as
\begin{equation}
\label{eq:olla}
\begin{split}
    \Delta_{OLLA} (t+1) = \Delta_{OLLA}(t) + \Delta_{up} . \mathds{1}[NACK] \\
    - \Delta_{down} . \mathds{1}[ACK]
    \end{split}
\end{equation}
\gls{mcs} selection is also updated for every new transmission, however, rank selection occur at the slow timescale of \gls{csi} reporting.
At each \gls{csi} reporting event k, the \gls{gNodeB} uses the current offset to compute the effective \gls{sinr} for every candidate rank $r$ using \eqref{eq:sinr}.
The selected rank $r^*$  then remains in use for all transmissions until the next \gls{csi} report arrives, during which the offset is continuously updated per \eqref{eq:olla} based on the \gls{harq} outcomes of the currently held rank.

The critical observation is that by the time the next \gls{csi} reporting event k+1 triggers rank selection, the offset $\Delta_{OLLA}$ may have accumulated many upward or downward updates, all reflecting the \gls{tbler} history of rank $r^*$ held since event k. 
Yet this accumulated offset is applied simultaneously to the effective \gls{sinr}s of all candidate ranks in \eqref{eq:sinr} at event k+1. 
The offset therefore encodes the sustained \gls{tbler} experience of one rank over an entire inter-\gls{csi} interval, but is used to predict throughput at every rank, a cross-rank contamination whose severity grows with the number of \gls{harq} feedback updates between consecutive \gls{csi} reports, and hence with traffic load.
Consequently, the network operates below its maximum achievable spectral efficiency, a particularly critical issue for \gls{xr} services, where such inefficiency directly reduces \gls{xr} capacity.

\subsection{\acrshort{moolla}}
To address the throughput prediction inaccuracy caused by cross-rank offset contamination, we propose the \gls{moolla} scheme.
In \gls{moolla}, each rank $r$ maintains an independent offset $\Delta_{OLLA}^r$, eliminating cross-rank contamination and enabling more accurate effective \gls{sinr} estimation $\hat\gamma_{eff}^{r}$ as follows

\begin{equation}
\label{eq:moolla}
    \hat\gamma_{eff}^{r} = \gamma^{r} - \Delta_{OLLA}^r
\end{equation}

The resulting $\hat\gamma_{eff}^{r}$ provides a more accurate throughput estimate reflecting only the accumulated \gls{tbler} history of rank $r$ itself. 
Rank and \gls{mcs} selection are then performed using $\hat\gamma_{eff}^{r}$ in place of $\gamma_{eff}^{r}$ in \eqref{eq:rankb}
In \gls{moolla}, \gls{harq} feedback is collected in the same manner as in the conventional single-offset \gls{olla} scheme.
For each transmission, only the \gls{olla} offset corresponding to the selected rank is updated based on the \gls{harq} feedback.
In this work, we assume same target \gls{tbler} for each \gls{olla} offset, so same step up and down sizes for each offset.
However, this scheme can be fine tuned for different target \gls{tbler}s per rank, by considering different step up or step down sizes per rank.

This approach improves spectral efficiency by enabling more accurate link adaptation.
The resulting spectral efficiency gains can be leveraged either to support a greater number of \gls{xr} users per cell or to accommodate other services such as \gls{embb}.

The \gls{moolla} procedure is summarized in Algorithm \ref{algo:olla1}.

\begin{algorithm}[t]
\small
\caption{\gls{moolla} with Independent Offsets per Rank}\label{algo:olla1}
\begin{algorithmic}[1]
\STATE \textbf{Initialization:} \\
Initialize \gls{olla} offsets  $\Delta_{OLLA}^{r}$ with the same value for all $r \in \{1,2,..\}$ \\
Set Step up size $\rightarrow$ $\Delta_{up}$ \\
Calculate down size $\rightarrow$ $\Delta_{down}$ based on  $TBLER^T$ and $\Delta_{up}$\\
\STATE \textbf{\gls{harq} feedback:} \\

$HF \rightarrow $ \gls{harq} feedback from Legacy \acrshort{ue} or \gls{thgr} \\

\IF {\gls{thgr}}
\STATE  $HF = HF_{J}$ \\
\ELSE
\STATE  $HF = HF_{X1}$ \\
\ENDIF
\STATE \textbf{\gls{olla} offsets Update:} \\
\STATE For the selected rank $r^*$ of the \gls{tb}
\IF {$HF \rightarrow$ ACK}
\STATE $\Delta_{OLLA}^{r^*} = \Delta_{OLLA}^{r^*} - \Delta_{down}$ \\
\ELSE
\STATE $\Delta_{OLLA}^{r^*} = \Delta_{OLLA}^{r^*} + \Delta_{up}$ \\

\ENDIF
\end{algorithmic}
\end{algorithm}

\section{Performance Evaluation}
\label{simrst}

\subsection{Evaluation Methodology}
We evaluate the performance of \gls{xr} \gls{thgr} against a baseline consisting of legacy \gls{xr} \gls{ue}s having direct link only, focusing on higher-rank transmissions.
Furthermore, the proposed \gls{moolla} scheme is evaluated for both legacy \gls{xr} \gls{ue}s and \gls{xr} \gls{thgr} and compared against the conventional single-offset \gls{olla}.
Initially, \gls{thgr} \gls{xr} capacity is evaluated assuming an ideal \gls{tl}, providing an upper-bound on achievable gains.
These upper-bound results are then compared against scenarios incorporating a realistic \gls{tl} model.

The evaluation is conducted using a \gls{sls} aligned with \gls{3gpp} evaluation guidelines for \gls{xr} \cite{bib2}.
The simulator features a comprehensive protocol stack and advanced \acrlong{rrm}, consistent with previous \gls{sls} studies \cite{bib2-our4} and the \gls{sls} tutorial \cite{bib3-8}.
Key simulation parameters are listed in Table \ref{tabpara}.

According to \gls{3gpp} \cite{bib2}, an \gls{xr} \gls{ue} is deemed happy if 99\% of packets are received correctly within the \gls{pdb}, while \gls{xr} capacity is defined as the maximum number of \gls{xr} \gls{ue}s per cell such that 90\% of them are happy.
Each simulation runs for 9~s after a 9~s warm-up period and is repeated over 10 independent drops per \gls{xr} load per cell.
This achieves a 99\% confidence level with a 1.1\% error margin for the \gls{xr} \gls{ue} happiness criterion while generating 540 packets(frames) per \gls{ue} after warm-up and a 2.1\% error margin for \gls{xr} capacity with 14 \gls{xr} \gls{ue}s per cell (12 cells x 10 drops x 14 \gls{ue}s/cell).

To reduce signaling overhead, only \gls{ue-x} is configured for periodic \gls{csi} reporting.
Prior work has shown that comparable performance is achieved regardless of whether \gls{csi} is reported by one or both \gls{ue}s (and best one is selected) when joint \gls{harq} processing and joint \gls{olla} algorithms are employed \cite{bib2-our4}.

\begin{figure}
\centerline{\includegraphics[width=8cm]{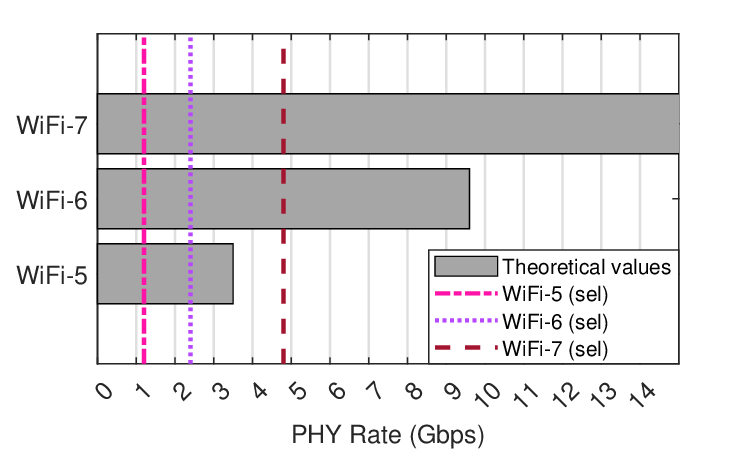}}
\caption{Throughput for \gls{wifi}-5 ( IEEE 802.11ac), \gls{wifi}-6 (IEEE 802.11ax) and \gls{wifi}-7 (IEEE 802.11be)}
\label{fig:wifi}
\end{figure}

Simulations are further repeated for different \gls{tl} configurations to evaluate \gls{xr} capacity under realistic conditions.
For the realistic \gls{tl}, per \gls{tb}(or soft-\gls{tb}/\gls{llr}) delay is computed using the model described in Section \ref{systemmodeltl} with representative \gls{wifi} PHY rates.
In this work, we select three representative operating points, approximately 1.2 Gbps, 2.4 Gbps, and 4.8 Gbps, to model practical \gls{tl}s, referred to hereafter as \gls{wifi}-5, \gls{wifi}-6 and \gls{wifi}-7, respectively. 
These labels refer to PHY rate operating points representative of each \gls{wifi} generation, not the full (theoretical) capabilities of each standard.
These PHY rates can be achieved using different \gls{wifi} generations \cite{bib7-17, bib7-18} through different parameter combinations as shown in Fig. \ref{fig:wifi}.
For instance, \gls{wifi}-7 (IEEE 802.11be) can theoretically exceeds 40 Gbps, while \gls{wifi}-6 (IEEE 802.11ax) supports peak rates up to 9.6 Gbps.
However, practically achievable rates are lower due to implementation and deployment constraints \cite{bib7-17, bib7-18, bib7-7}.
For \gls{llr} forwarding, a 5-bit quantization is assumed, which provides near lossless performance in iterative decoding compared to unquantized \gls{llr} \cite{bib2-8}.

\begin{table}[h]
\caption{Simulation Parameters}
\begin{center}
\resizebox{0.85\columnwidth}{!}{
\begin{tabular}{|c|c|}
\hline
\textbf{Parameter} & \textbf{Setting} \\
\hline
Deployment & Indoor Hotspot (120m $\times$ 50m) \cite{bib13}\\
\hline
Layout & 12 cells\\
\hline
Inter-site Distance & 20m\\
\hline
TTI length & 14 \acrshort{ofdm} symbols\\
\hline
\multirow{1}{*}{\acrshort{tdd} Frame Structure} & DDDSU\\
\hline
Bandwidth & 100 MHz \\
\hline
Carrier Frequency & 4 GHz \\
\hline
\acrlong{s-c-s} & 30 kHz \\
\hline
Modulation & QPSK to 256QAM  \\
\hline
\gls{xr} frame rate & 60 fps \\
\hline
\gls{xr} random jitter & $\mathcal{TN}(0,2,-4,4)$ ms\\
\hline
\gls{xr} frame size (45 Mbps) & $\mathcal{TN}(93,10,46,141)$ kB\\
\hline
\gls{gNodeB} height  & 3m \\ 
\hline
\gls{gNodeB} power & 31 dBm \\
\hline
\multirow{2}{*}{\gls{gNodeB} antenna} & 1 panel with 32 elements\\
& (4 $\times$ 4 and 2 polarization) \\
\hline
\gls{gNodeB} \acrshort{tx} processing delay & 2.75 \acrshort{ofdm} symbols \\
\hline
\gls{ue} height &  1.5m \\
\hline
\multirow{2}{*}{\gls{ue} antenna} & 4 omni-directional antennas \\
 & (2 $\times$ 1 and 2 polarization)\\
\hline

\gls{ue} receiver & \acrshort{mmseirc} \\
\hline
\gls{ue} \acrshort{rx} processing delay & 6 \acrshort{ofdm} symbols\\
\hline
Rank & Up to 4 \\
\hline
Scheduler & Proportional Fair \\
\hline
\acrshort{cqi} Measurement & Periodic every 2 ms\\
\hline
\acrshort{cqi} Reporting Delay & 2 ms \\
\hline
Channel Estimation & Ideal \\
\hline
\gls{harq} combining method & Chase soft combining \\
\hline
Target \gls{bler} & 10\%\\
\hline
Intra \gls{thgr} distance & 1m \\
\hline
\multicolumn{2}{|c|}{\gls{wifi} Parameters} \\
\hline
$CW_{min}$ & 15\\
\hline
$CW_{max}$ & 1023 \\
\hline
$N_{max}$ & 3\\
\hline
$\sigma$ & 9 $\mu$ s\\
\hline
$SIFS$ & 16 $\mu$ s\\
\hline
$DIFS$ & 34 $\mu$ s\\
\hline
$T_{RTS}$ & 27 $\mu$ s\\
\hline
$T_{CTS}$ & 19 $\mu$ s\\
\hline
$T_{PHY}$ & 120 $\mu$ s\\
\hline
$L_H$ & 36 B \\
\hline
$p_c$ & 10\% \\
\hline
\end{tabular}

}
\label{tabpara}
\end{center}
\end{table}

\subsection{Simulation Results}

\subsubsection{Ideal Tethering Link}
\begin{figure}[t]
    \centering
    \begin{subfigure}[b]{0.45\columnwidth}
        \centering
        \includegraphics[width=4.5cm]{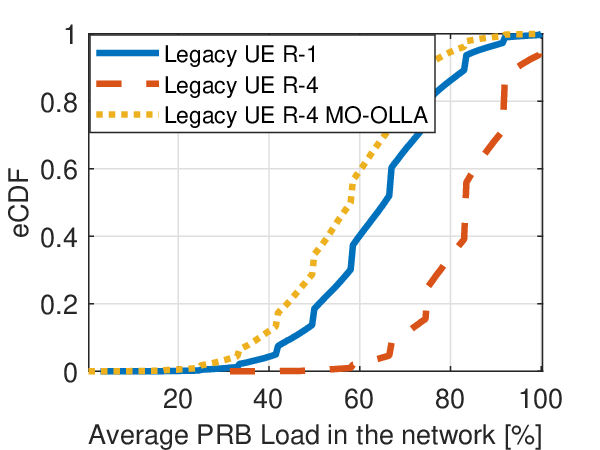}
        \caption{5 Legacy UEs/cell}
    \end{subfigure}
    \hfill
    \begin{subfigure}[b]{0.45\columnwidth}
        \centering
        \includegraphics[width=4.5cm]{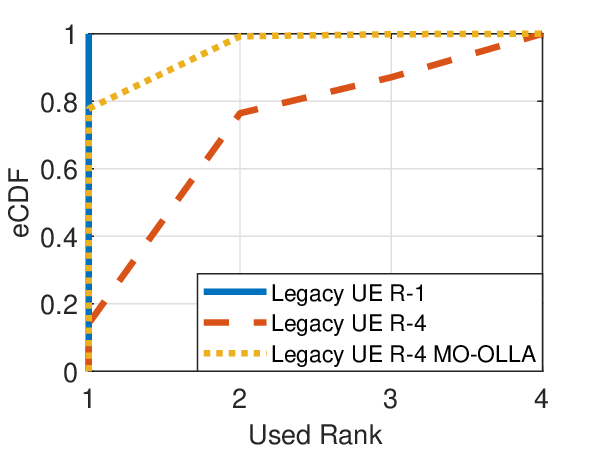}
        \caption{5 Legacy UEs/cell}
        \label{fig:usedrankue}
    \end{subfigure}

    \begin{subfigure}[b]{0.45\columnwidth}
        \centering
        \includegraphics[width=4.5cm]{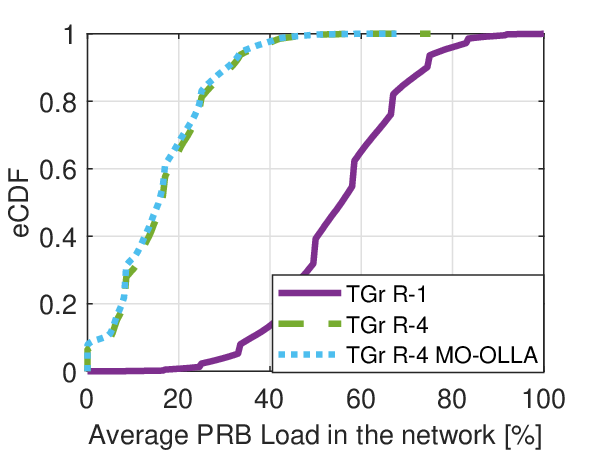}
        \caption{5 TGrs/cell with Ideal TL}
    \end{subfigure}
    \hfill
    \begin{subfigure}[b]{0.45\columnwidth}
        \centering
        \includegraphics[width=4.5cm]{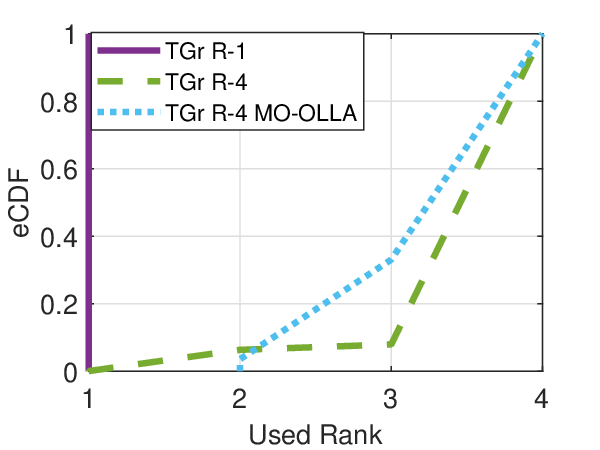}
        \caption{5 TGrs/cell with Ideal TL}
    \end{subfigure}

    \begin{subfigure}[b]{0.45\columnwidth}
        \centering
        \includegraphics[width=4.5cm]{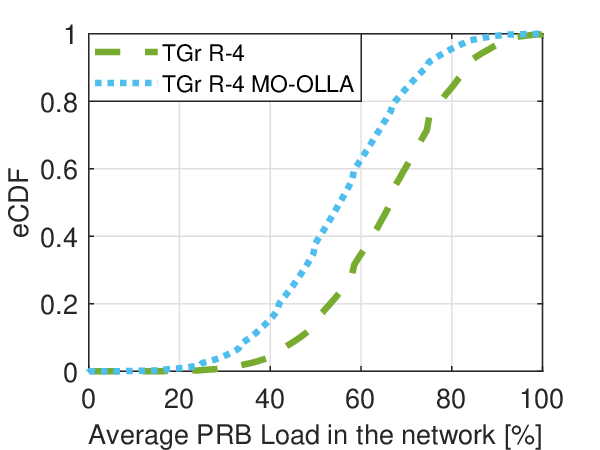}
        \caption{12 TGrs/cell with Ideal TL}
    \end{subfigure}
    \hfill
    \begin{subfigure}[b]{0.45\columnwidth}
        \centering
        \includegraphics[width=4.5cm]{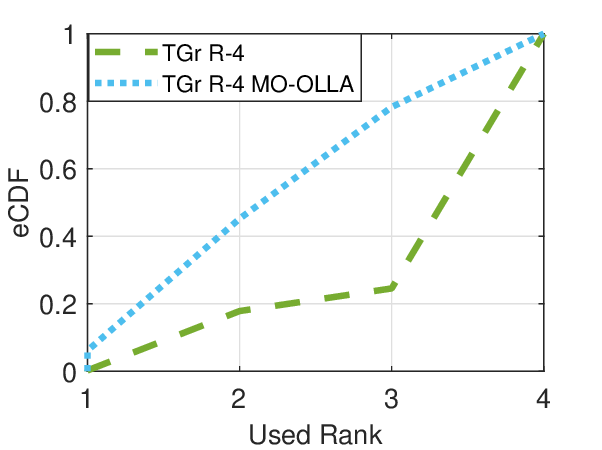}
        \caption{12 TGrs/cell with Ideal TL}
    \end{subfigure}
    
    \caption{Average \acrshort{prb} utilization and used rank in the network.}
    \label{fig:2x2prbrank}
\end{figure}

Fig. \ref{fig:2x2prbrank} presents the \gls{ecdf} of the average \gls{prb} utilization in the network, along with the distribution of the selected transmission rank, for both legacy \gls{ue}s and \gls{thgr}s.
Results are presented for three configurations R-1 (rank-1), R-4 (up to rank-4 with single \gls{olla} offset) and R-4 \gls{moolla} (up to rank-4 with \gls{moolla}).
For legacy \gls{ue}s, R-4  results in higher \gls{prb} utilization than R-1, reflecting the impact of suboptimal rank selection.
In contrast, \gls{thgr} do not exhibit this behavior owing to the cooperative gains.
\gls{thgr} improve decoding reliability and reduce \gls{ici} by operating at higher spectral efficiency (i.e., higher-rank and \gls{mcs}) thereby requiring fewer \gls{prb}s.
Nevertheless, even for \gls{thgr}, a single \gls{olla} offset does not guarantee optimal rank selection. 
Enabling \gls{moolla} improves rank selection accuracy for both legacy \gls{ue}s and \gls{thgr}s, resulting in reduced \gls{prb} utilization under R-4 \gls{moolla} compared to R-4. 
For \gls{thgr}, this gain is more pronounced at higher load (12 \gls{thgr}s per cell) than at lower load (5 \gls{thgr}s per cell), since \gls{ici} depends jointly on the number of active users and their resource utilization efficiency.
\gls{thgr} tends to operate at higher-ranks than legacy \gls{ue}s due to cooperation gains, which are effectively captured by the joint \gls{olla} algorithm.
For example, with 5 users (legacy \gls{ue}s/\gls{thgr}s) per cell and \gls{moolla}, legacy \gls{ue}s select rank-4 in fewer than $1\%$ of transmissions, whereas \gls{thgr} select rank-4 in approximately $67\%$ of transmissions. 
Consequently, at the 50th percentile with 5 users per cell and \gls{moolla}, \gls{prb} utilization is 58\% for legacy \gls{ue}s compared to only 16\% for \gls{thgr}s.
Results for legacy \gls{ue}s with 12 users per cell are omitted, as this load level cannot be supported due to excessive \gls{prb} utilization.
Similarly, \gls{thgr} results under rank-1 transmission with 12 users per cell are omitted for same reason \cite{bib2-our4}.
  
\begin{figure}[t]
    \centering
    \begin{subfigure}[b]{0.45\textwidth}
        \centering
        \includegraphics[width=8.5cm]{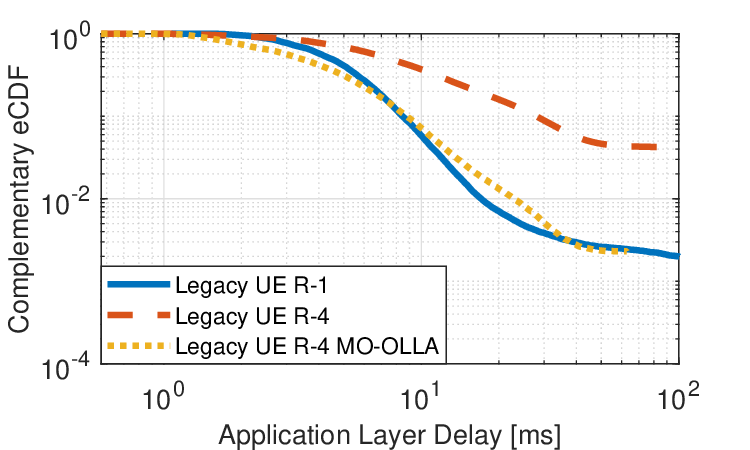}
        \caption{5 Legacy UEs/cell}
    \end{subfigure}
    \hfill
    \begin{subfigure}[b]{0.45\textwidth}
        \centering
        \includegraphics[width=8.5cm]{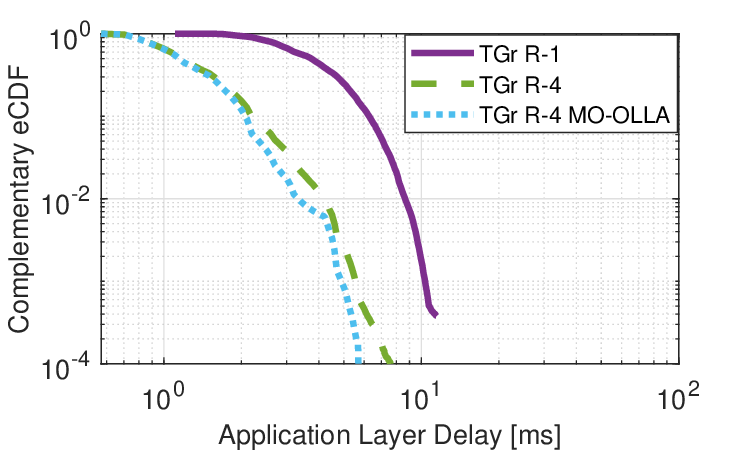}
        \caption{5 TGrs/cell with Ideal TL}
    \end{subfigure}
    \begin{subfigure}[b]{0.45\textwidth}
        \centering
        \includegraphics[width=8.5cm]{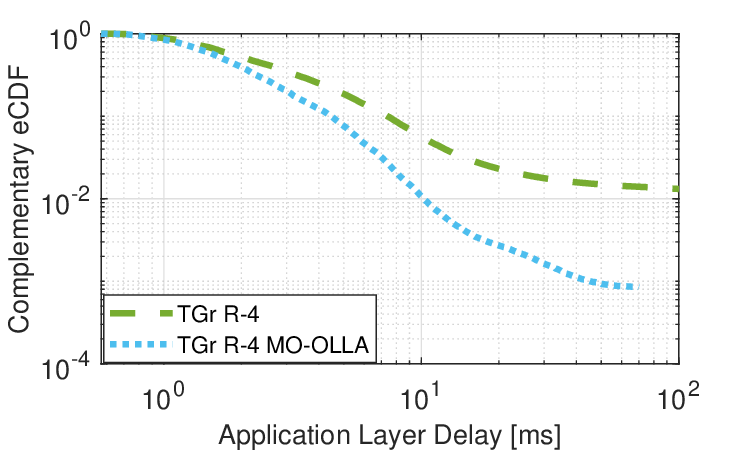}
        \caption{12 TGrs/cell with Ideal TL}
    \end{subfigure}

    \caption{Application Layer Delay for legacy UEs and TGrs with Ideal TL.}
    \label{fig:delay}
\end{figure}

Fig. \ref{fig:delay} shows the complementary \gls{ecdf} of application layer delay for both legacy \gls{ue}s and \gls{thgr}.
\gls{moolla} reduces delay compared to single-offset \gls{olla} for both legacy \gls{ue}s and \gls{thgr}s.
For \gls{thgr}, \gls{moolla} outperforms R-1, whereas for legacy \gls{ue}s, the improvement is not uniform across all transmissions and/or users. 
This non-uniformity arises because higher-rank transmissions can degrade the interference suppression capability of the \gls{mmseirc} receiver, depending on the available degrees of freedom (i.e., number of receive antennas relative to the number of interferers) \cite{bib7-14}.
As a result, some legacy \gls{ue}s that remain at rank-1 under R-4 \gls{moolla} experience degraded delay performance relative to the R-1 case.
In contrast, most \gls{thgr}s operate at higher-ranks due to cooperation gains, avoiding this issue.
The difference in delay performance between R-4 and R-4 \gls{moolla} is smaller for \gls{thgr}s with lower load (5 \gls{thgr}s per cell) than at higher load (12 \gls{thgr}s per cell).
Overall, \gls{thgr}s achieve significantly lower application layer delay compared to legacy \gls{ue}s. 

\begin{figure}[t]
    \centering
    \begin{subfigure}[b]{0.45\columnwidth}
        \centering
        \includegraphics[width=4.5cm]{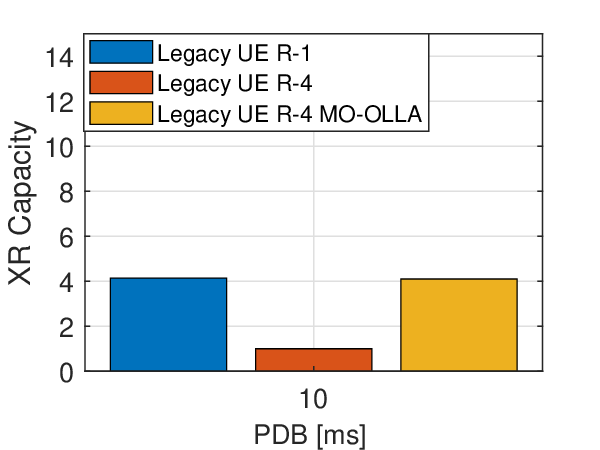}
        \caption{Legacy UE (10 ms)}
    \end{subfigure}
    \hfill
    \begin{subfigure}[b]{0.45\columnwidth}
        \centering
        \includegraphics[width=4.5cm]{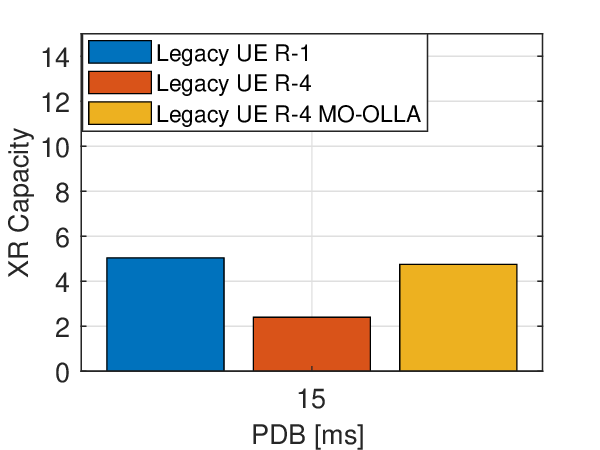}
        \caption{Legacy UE (15 ms)}
    \end{subfigure}

    \begin{subfigure}[b]{0.45\columnwidth}
        \centering
        \includegraphics[width=4.5cm]{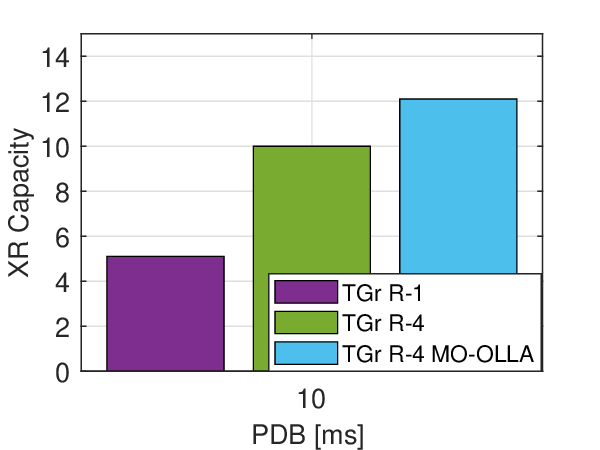}
        \caption{TGr with Ideal TL (10 ms)}
    \end{subfigure}
    \hfill
    \begin{subfigure}[b]{0.45\columnwidth}
        \centering
        \includegraphics[width=4.5cm]{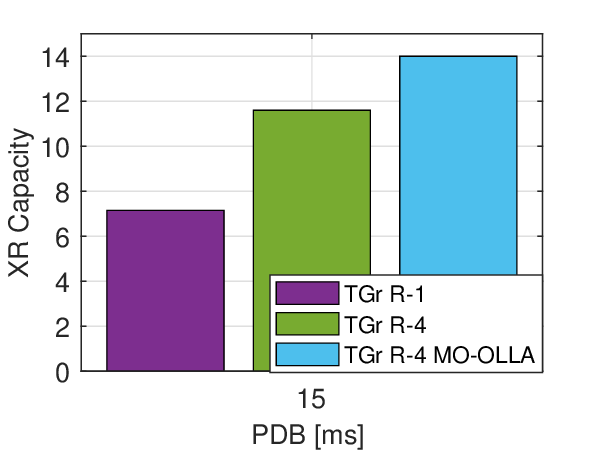}
        \caption{TGr with Ideal TL (15 ms)}
    \end{subfigure}
    
    \caption{XR Capacity of the network with legacy XR UEs and XR TGrs.}
    \label{fig:2x2cap}
\end{figure}

Fig. \ref{fig:2x2cap} presents the \gls{xr} capacity for both legacy \gls{ue}s and \gls{thgr}s under two \gls{pdb} constraints, i.e., 10 ms and 15 ms.
For legacy \gls{ue}s, enabling higher-rank transmission with a single-offset \gls{olla} reduces \gls{xr} capacity due to suboptimal rank selection.
\gls{moolla} improves performance for  legacy \gls{ue}s but still yields slightly lower capacity than rank-1 operation attribute to the impact of higher-rank transmissions on \gls{mmseirc} interference suppression.
In contrast, \gls{thgr} benefits substantially from higher-rank transmissions, both R-4 and R-4 \gls{moolla} outperform R-1, with \gls{moolla} yielding the highest \gls{xr} capacity.
For example, under a 10 ms \gls{pdb} constraint, \gls{thgr} with rank-4 and \gls{moolla} supports up to 12 \gls{xr} users per cell compared to 10 users with single-offset \gls{olla}. 
The capacity gains of \gls{thgr}s over legacy \gls{ue}s are substantial.
For instance, under 10 ms and 15 ms \gls{pdb} constraints, legacy \gls{ue}s achieve maximum \gls{xr} capacities (either with R-1 or R-4 \gls{moolla}) of 4.1 and 5.04 \gls{xr} \gls{ue}s per cell, respectively, whereas \gls{thgr}s achieve 12.1 and 14 \gls{xr} \gls{ue}s per cell. 
The gains stem from the ability of \gls{thgr}s to operate at higher spectral efficiency (higher \gls{mcs} and rank due to enhanced decoding reliability) for a given \gls{tbler} and \gls{sinr}, thereby reducing \gls{ici} and further improving \gls{sinr} in a compound effect enabled by the joint \gls{olla} and joint \gls{harq} algorithms.
\gls{moolla} yields a 20\% improvement in \gls{thgr} performance.
Combined with \gls{moolla}, this results in upper-bound \gls{xr} capacity gains for \gls{thgr} in the range of 180-200\% relative to legacy \gls{ue}s under ideal \gls{tl} conditions.

\subsubsection{Realistic Tethering Link}
We next evaluate the performance of \gls{thgr}s under realistic \gls{tl} conditions, comparing \gls{thgr} with an ideal \gls{tl} (TGr-TL-Ideal) against \gls{thgr}s employing \gls{wifi}-5, \gls{wifi}-6 and \gls{wifi}-7 as \gls{tl}s.
In all cases, up to rank 4 transmissions and \gls{moolla} are used.

\begin{figure}[htbp]
    \centering
        \centering
        \includegraphics[width=8cm]{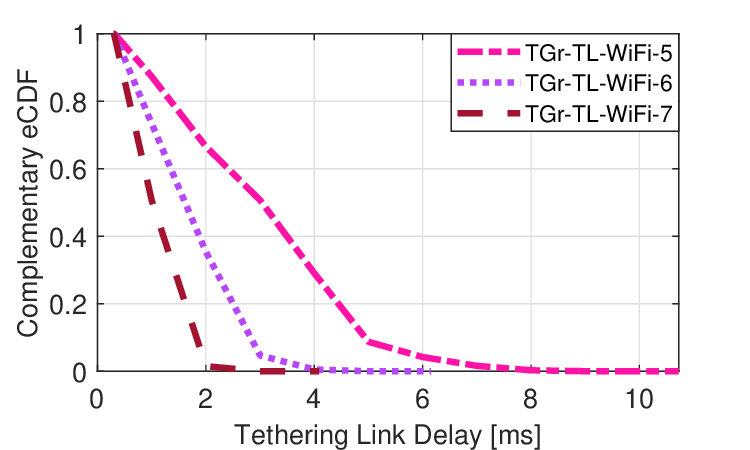}

    \caption{TL Delay with different WiFi (PHY rate).}
    \label{fig:tldelay}
\end{figure}

Fig. \ref{fig:tldelay} presents the complementary \gls{ecdf} of \gls{tl} delay for 12 \gls{thgr}s per cell.
As expected, lower PHY rates results in higher delays, with the tail of the distribution corresponding to above average \gls{tb} sizes.
The mean delays are 2.98 ms, 1.64 ms and 1 ms with \gls{wifi}-5, \gls{wifi}-6 and \gls{wifi}-7, respectively, while the maximum observed delays are 10.73 ms, 4.1 ms and 6.15 ms.

\begin{figure}[t]
    \centering
    \begin{subfigure}[b]{0.48\columnwidth}
        \centering
        \includegraphics[width=4.8cm]{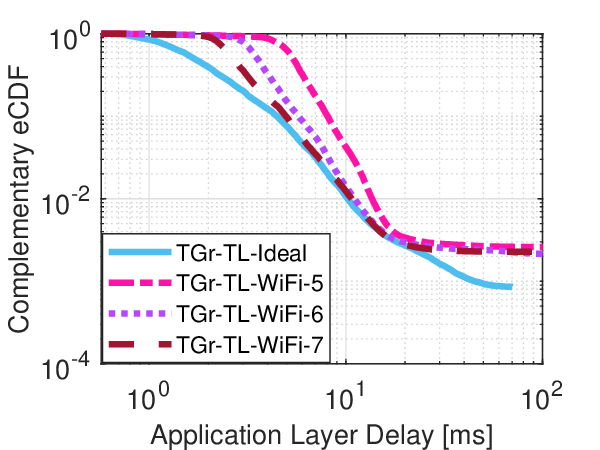}
    \end{subfigure}
    \hfill
    \begin{subfigure}[b]{0.48\columnwidth}
        \centering
        \includegraphics[width=4.8cm]{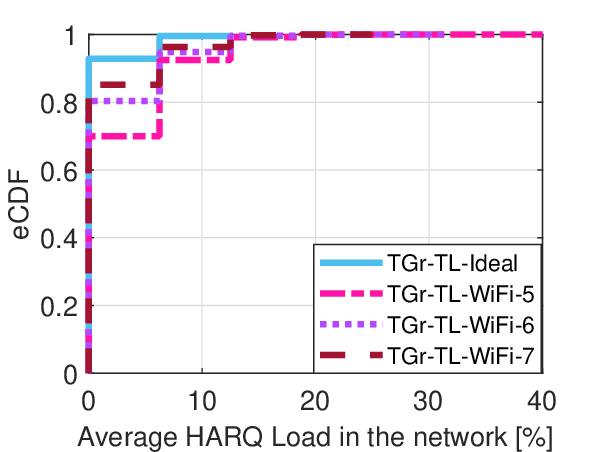}
    \end{subfigure}
    
    \caption{Effects of Realistic TL delay on cellular network}
    \label{fig:tlimpact}
\end{figure}

\gls{tl} delay impacts both application layer delay and \gls{harq} resource utilization.
Delayed \gls{harq} feedback (i.e., the second \gls{harq} from \acrshort{ue-x}) prolongs \gls{saw} channel occupancy referred here as \gls{harq} load in the network, defined as the ratio of number of SAW channels occupied / total number of SAW channels.
Fig. \ref{fig:tlimpact} illustrates the application layer delay and average \gls{harq} load under realistic \gls{tl}.
The \gls{harq} load remains limited, even at the 100th percentile, \gls{harq} load does not exceed 43.75\%, 31.25\% and 25\% for \gls{wifi}-5, \gls{wifi}-6 and \gls{wifi}-7 links, respectively, compared to 12.5\% for ideal links.
\gls{harq} resource exhaustion is therefore not observed under the given conditions.
However, application layer delay is considerably affected, at the 99\% percentile, application layer delay increase from 10.22 ms (ideal) to 13.6 ms, 11.07 ms and 10.71 ms for \gls{wifi}-5, \gls{wifi}-6 and \gls{wifi}-7, respectively.

\begin{figure}[t]
    \centering
    \begin{subfigure}[b]{0.48\columnwidth}
        \centering
        \includegraphics[width=4.8cm]{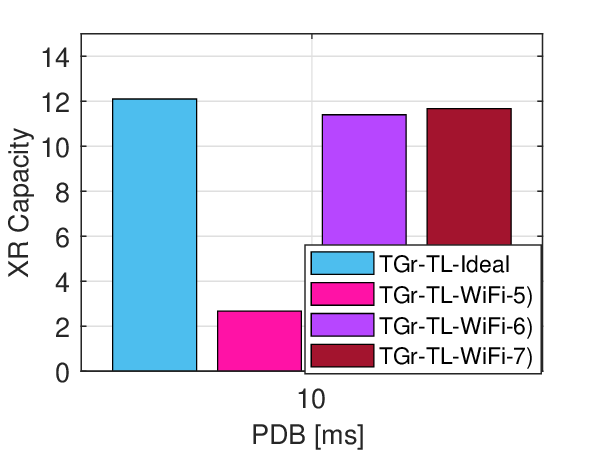}
    \end{subfigure}
    \hfill
    \begin{subfigure}[b]{0.48\columnwidth}
        \centering
        \includegraphics[width=4.8cm]{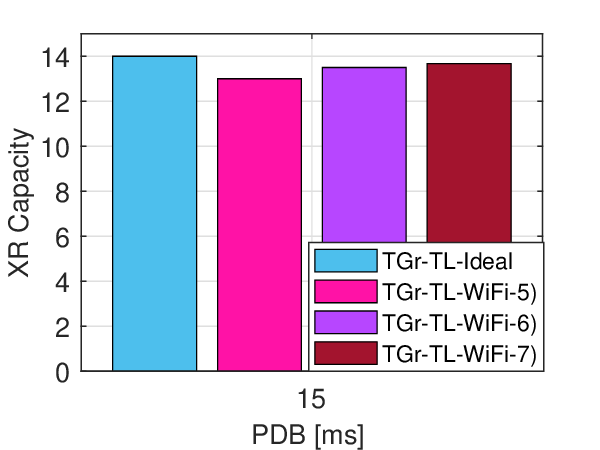}
    \end{subfigure}
    
    \caption{XR Capacity with \gls{thgr} having Realistic TL Delay with max rank-4 and \gls{moolla}.}
    \label{fig:xrcapwtl}
\end{figure}
Fig. \ref{fig:xrcapwtl} compares \gls{xr} capacity under realistic \gls{tl}s with the ideal case.
Under a 15 ms \gls{pdb} constraint, performances degradation with realistic \gls{tl} is minimal, \gls{thgr} with \gls{wifi}-5 \gls{tl} supports only one fewer \gls{xr} \gls{ue} per cell than the ideal case, while \gls{thgr} with \gls{wifi}-6 and \gls{wifi}-7 achieve \gls{xr} capacities of 13.5 and 13.7 \gls{xr} \gls{ue}s per cell, respectively, compared to the ideal maximum of 14.
Under the more stringent 10 ms \gls{pdb} constraint, \gls{thgr} performance degrades significantly with \gls{wifi}-5, falling even below legacy \gls{ue} performance.
This occurs because \gls{tb}s that fail decoding even after soft combining incur retransmissions triggered by delayed \gls{harq} feedback, resulting in worst case delays that exceed those of legacy \gls{ue}s.   
In contrast, \gls{wifi}-6 and \gls{wifi}-7 maintain strong performance, supporting nearly the same \gls{xr} capacity as the ideal case and provide 165-180\% \gls{xr} capacity gains relative to legacy \gls{ue}s.


The results collectively establish that \gls{thgr} based cooperation delivers substantial and practically realizable \gls{xr} capacity gains over legacy single-link \gls{ue}s. 
Under ideal tethering conditions, combining higher-rank transmission with \gls{moolla} yields \gls{xr} capacity gains of 180-200\% relative to legacy \gls{ue}s, gains that arise from the compound effect of enhanced decoding reliability, reduced \gls{ici}, and more accurate rank-aware link adaptation enabled by the simple joint \gls{olla}, joint \gls{harq}, and \gls{moolla} algorithms. 
\gls{moolla} scheme enables optimal rank selection with minimal additional complexity, requiring only per-rank \gls{olla} offsets without modifications to existing standards.
A key question for practical deployment is how much of this gain survives under realistic \gls{tl} constraints. 
The results provide a clear and encouraging answer: with PHY rates of approximately 2.4 Gbps and above, \gls{thgr}s provide 165-180\% \gls{xr} capacity gains relative to legacy \gls{ue}s, with only minor degradation relative to the ideal case. 
This robustness stems from the fact that, at these PHY rates, mean \gls{tl} delays of 1-1.64 ms keep the delayed \gls{harq} feedback from \acrshort{ue-x} well within the limit, thereby avoiding the worst-case retransmission chains that drive application layer delay beyond the \gls{pdb}. 
Importantly, \gls{harq} channel occupancy remains well below saturation across all evaluated \gls{wifi} configurations, even at the 100th percentile, 
The one exception is the combination of \gls{wifi}-5 equivalent tethering (PHY rate 1.2 Gbps) with a 10 ms \gls{pdb} constraint, where \gls{tl} delays can trigger retransmissions that push end-to-end delay beyond the \gls{pdb}, causing \gls{thgr} performance to fall below that of legacy \gls{ue}s. 
However, the \gls{thgr}s with \gls{tl} PHY rates of 2.4 Gbps or above preserve the cooperative advantage effectively and providing gains approximately near to ideal case. 
Taken together, these findings confirm the practical viability of \gls{thgr} based cooperation for \gls{xr} capacity enhancement for future \gls{xr} deployments.

\section{Conclusion}

\label{conc}


This paper investigated the performance of multi-connected \gls{xr} devices enabled through \gls{thgr} under higher-rank \gls{ptm} transmission, with a focus on link adaptation accuracy and practical \gls{tl} constraints. 
The results demonstrate that, beyond reliability improvements, \gls{thgr} operation can effectively leverage spatial multiplexing better than conventional single-link \gls{ue}s. 
This leads to substantially more efficient resource utilization and a significant increase in \gls{xr} capacity.
The proposed \gls{moolla} framework addresses a fundamental limitation of conventional single-offset \gls{olla}, its inability to provide accurate \gls{sinr} correction across ranks.
By maintaining independent offsets per rank, \gls{moolla} improves rank selection accuracy and yields up to 20\% improvement in \gls{thgr} \gls{xr} capacity with no modifications to existing standards, making it a practical enhancement for deployment.
System-level evaluations confirm that \gls{thgr} achieves substantial \gls{xr} capacity gains of 180–200\% over single-link \gls{xr} \gls{ue}s under ideal \gls{tl}.
Analysis of the realistic \gls{tl} model reveals that tethering induced delays have a limited impact on \gls{harq} resource utilization, and do not significantly degrade system performance provided the \gls{tl} offers sufficient PHY rate. 
Specifically, \gls{tl}s operating at Wi-Fi-6 or Wi-Fi-7 equivalent PHY rates (~2.4 Gbps and above) are shown to provide 165–180\% \gls{xr} capacity gains relative to legacy \gls{ue}s, confirming that the cooperation benefits of \gls{thgr} are not merely theoretical, 90\% of the \gls{thgr} gains remain intact under realistic deployment conditions. 
These findings highlight the practical viability of \gls{thgr} based cooperation as an effective mechanism for supporting higher \gls{xr} capacity within existing cellular network resources, and provide actionable guidance for \gls{tl} selection and device pairing in future \gls{xr} deployments.

\begin{IEEEbiography}[{\includegraphics[width=1in,height=1.25in,clip,keepaspectratio]{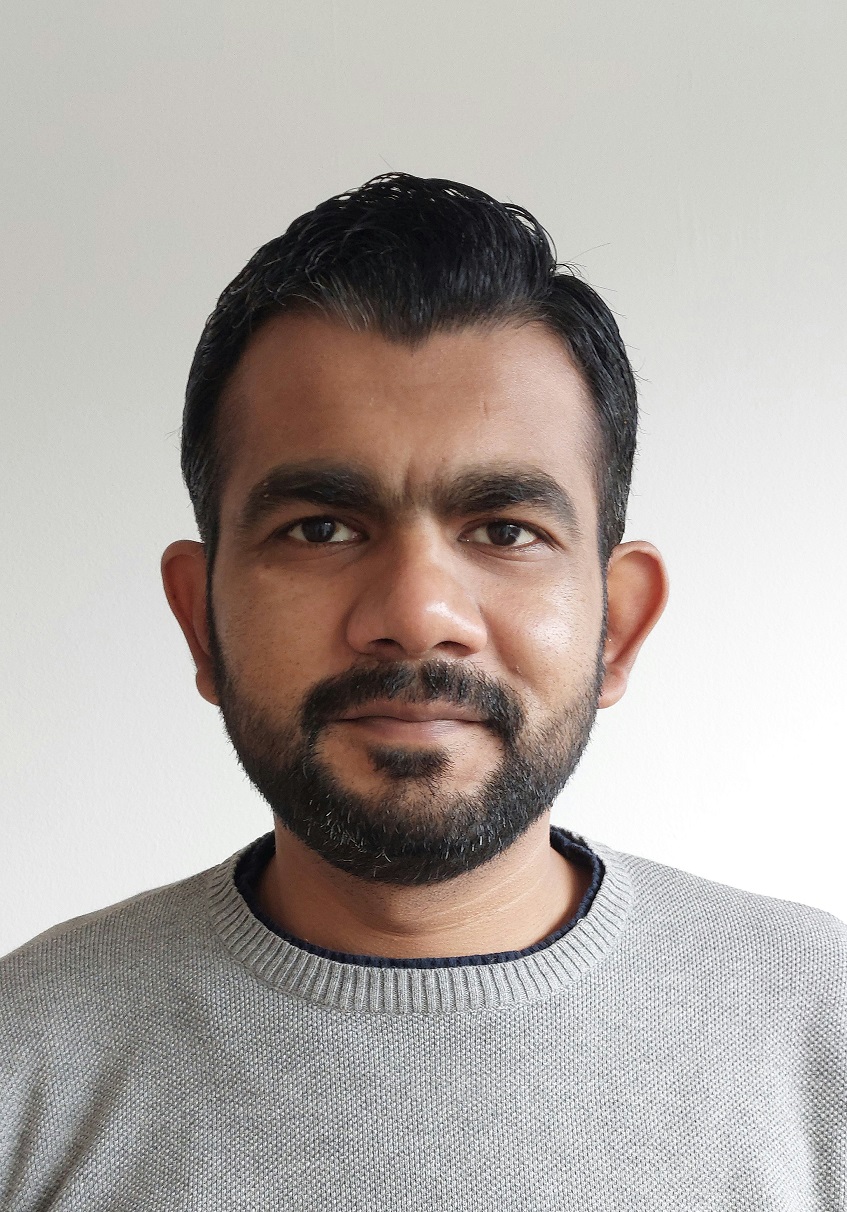}}]{\textbf{MUHAMMAD AHSEN }}  (Graduate Student Member, IEEE) received the B.E. degree in Electrical Engineering in 2013 and the M.S. degree(Hons.) in Electrical Engineering in 2016 from the National University of Sciences and Technology (NUST), Islamabad, Pakistan. He is currently pursuing the Ph.D. degree with the Department of Electronics Systems, Aalborg University, Denmark, in collaboration with Nokia Standard, Aalborg, Denmark. His research interest includes extended reality communications, 5G/6G radio resource management, cooperative communication and time-sensitive wireless communication.  
\end{IEEEbiography}

\begin{IEEEbiography}[{\includegraphics[width=1in,height=1.25in,clip,keepaspectratio]{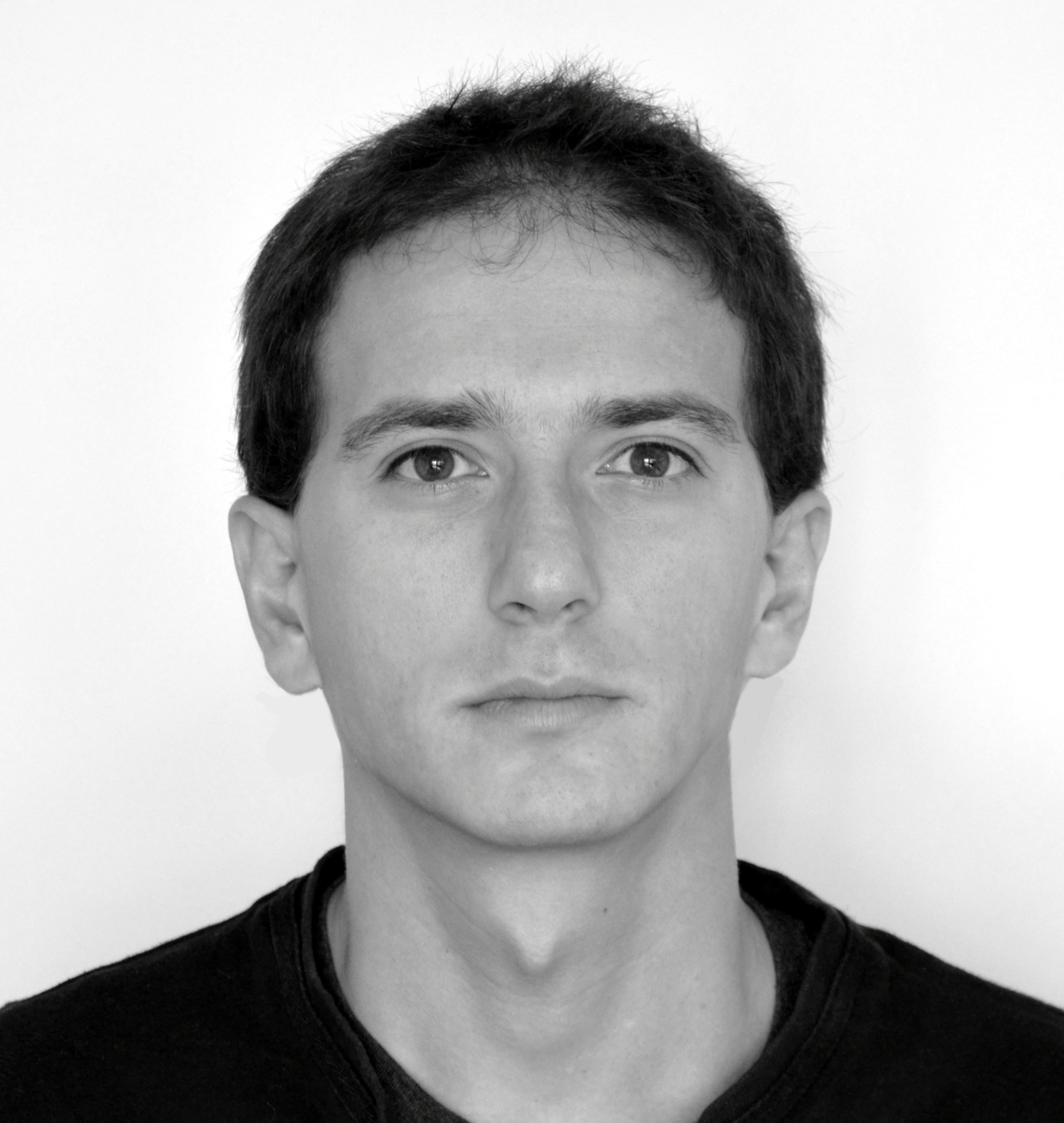}}]{\textbf{BOYAN YANAKIEV }} received the B.Sc. degree in physics from Sofia University, Bulgaria, in 2006, and the M.Sc. and Ph.D. degrees from Aalborg University, Aalborg, Denmark, in 2008 and 2011, respectively. He is currently with Nokia Standardization Research, Denmark, focusing on 5G advanced and 6G topics. His current work is in the area of XR optimization in RAN  
\end{IEEEbiography}

\begin{IEEEbiography}[{\includegraphics[width=1in,height=1.25in,clip,keepaspectratio]{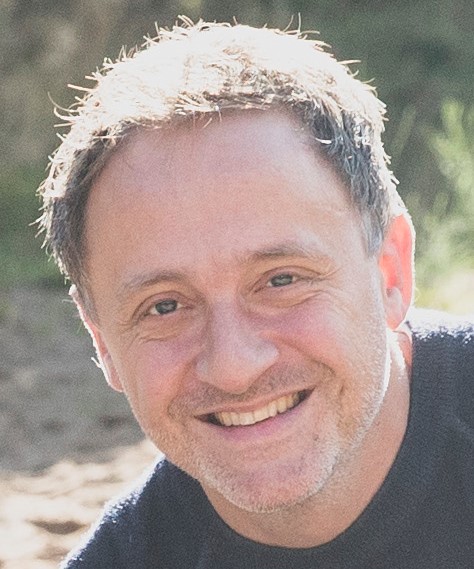}}]{\textbf{CLAUDIO ROSA }} received his M.Sc. E.E. and Ph.D. degrees in 2000 and 2005, respectively, from Aalborg University. In 2003 he also received an M.Sc.E.E. degree in telecommunication engineering from Politecnico di Milano, Italy. Since he joined Nokia in 2005, he contributed to standardization of 4G and 5G systems working on uplink power control and radio resource management, carrier aggregation, dual connectivity, and unlicensed spectrum operation. His current research interests also include user plane protocol design for 6G, flexible duplexing and radio enablers for eXtended reality applications. He has filed more than 200 patent applications, holds more than 50 granted patents, and has co-authored more than 50 scientific publications.   
\end{IEEEbiography}

\begin{IEEEbiography}[{\includegraphics[width=1in,height=1.2in,clip]{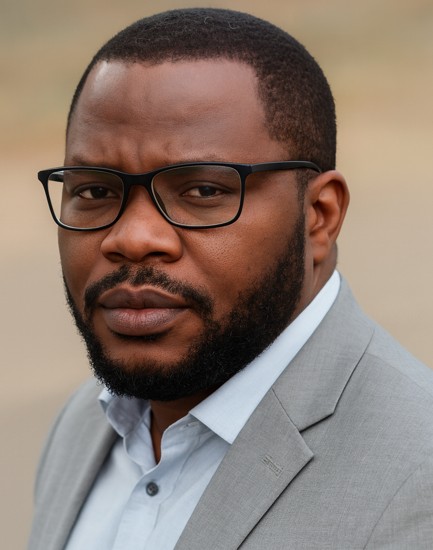}}]{\textbf{RAMONI ADEOGUN }} (Senior Member, IEEE) received the B.Eng. degree in electrical and computer engineering from the Federal University of Technology, Minna, Nigeria, and the Ph.D. degree in electronic and computer systems engineering from the Victoria University of Wellington, New Zealand in 2007 and 2015, respectively. He is currently an Associate Professor and leader of the AI for Communications group at Aalborg University, Denmark. Prior to joining Aalborg University, he held varios academic and industry research positions at University of Cape Town, South Africa, Odua Telecoms Ltd., and the National Space Research and Development Agency, Nigeria. His research interests include signal processing, machine learning and AI for PHY MAC and RRM. He has co-authored over 70 peer-reviewed publications. 
\end{IEEEbiography}

\end{document}